\documentstyle[galley,epsf,booktabs]{mn}
\newif\ifAMStwofonts

\ifoldfss
  \ifCUPmtlplainloaded \else
    \NewTextAlphabet{textbfit} {cmbxti10} {}
    \NewTextAlphabet{textbfss} {cmssbx10} {}
    \NewMathAlphabet{mathbfit} {cmbxti10} {} 
    \NewMathAlphabet{mathbfss} {cmssbx10} {} 
  \fi
  \ifAMStwofonts
    \ifCUPmtlplainloaded \else
      \NewSymbolFont{upmath} {eurm10}
      \NewSymbolFont{AMSa} {msam10}
      \NewMathSymbol{\upi}     {0}{upmath}{19}
      \NewMathSymbol{\umu}     {0}{upmath}{16}
      \NewMathSymbol{\upartial}{0}{upmath}{40}
      \NewMathSymbol{\leqslant}{3}{AMSa}{36}
      \NewMathSymbol{\geqslant}{3}{AMSa}{3E}

    \fi
  \fi
\fi 

\ifnfssone
  \newmathalphabet{\mathit}
  \0addtoversion{normal}{\mathit}{cmr}{m}{it}
  \addtoversion{bold}{\mathit}{cmr}{bx}{it}
  \newmathalphabet{\mathbfit} 
  \addtoversion{normal}{\mathbfit}{cmr}{bx}{it}
  \addtoversion{bold}{\mathbfit}{cmr}{bx}{it}
  \newmathalphabet{\mathbfss} 
  \addtoversion{normal}{\mathbfss}{cmss}{bx}{n}
  \addtoversion{bold}{\mathbfss}{cmss}{bx}{n}
  \ifAMStwofonts
    \ifCUPmtlplainloaded \else
      \UseAMStwoboldmath
      \makeatletter
      \new@mathgroup\upmath@group
      \define@mathgroup\mv@normal\upmath@group{eur}{m}{n}
      \define@mathgroup\mv@bold\upmath@group{eur}{b}{n}
      \edef\UPM{\hexnumber\upmath@group}
      \new@mathgroup\amsa@group
      \define@mathgroup\mv@normal\amsa@group{msa}{m}{n}
      \define@mathgroup\mv@bold\amsa@group{msa}{m}{n}
      \edef\AMSa{\hexnumber\amsa@group}
      \makeatother
      \mathchardef\upi="0\UPM19
      \mathchardef\umu="0\UPM16
      \mathchardef\upartial="0\UPM40
      \mathchardef\leqslant="3\AMSa36
      \mathchardef\geqslant="3\AMSa3E
    \fi
  \fi
\fi 

\ifnfsstwo
  \DeclareMathAlphabet{\mathbfit}{OT1}{cmr}{bx}{it}
  \SetMathAlphabet\mathbfit{bold}{OT1}{cmr}{bx}{it}
  \DeclareMathAlphabet{\mathbfss}{OT1}{cmss}{bx}{n}
  \SetMathAlphabet\mathbfss{bold}{OT1}{cmss}{bx}{n}
  \ifAMStwofonts
    \ifCUPmtlplainloaded \else
      \DeclareSymbolFont{UPM}{U}{eur}{m}{n}
      \SetSymbolFont{UPM}{bold}{U}{eur}{b}{n}
      \DeclareSymbolFont{AMSa}{U}{msa}{m}{n}
      \DeclareMathSymbol{\upi}{0}{UPM}{"19}
      \DeclareMathSymbol{\umu}{0}{UPM}{"16}
      \DeclareMathSymbol{\upartial}{0}{UPM}{"40}
      \DeclareMathSymbol{\leqslant}{3}{AMSa}{"36}
      \DeclareMathSymbol{\geqslant}{3}{AMSa}{"3E}
    \fi
  \fi
\fi 

\ifCUPmtlplainloaded \else
  \ifAMStwofonts \else 
    \def\upi{\pi}
    \def\umu{\mu}
    \def\upartial{\partial}
  \fi
\fi

\begin{document}
\title[MidIR variations of RCB Stars]
   {Mid-Infrared Variations of  R Coronae Borealis  Stars}

\author[N. Kameswara Rao, and David L. Lambert]
       {N. Kameswara Rao$^1$$^,$$^2$, \& David L. Lambert$^2$ \\
       $^1$Indian Institute of Astrophysics, Bangalore 560034, India\\
       $^2$The W.J. McDonald Observatory and Department of Astronomy, The University of Texas at Austin, Austin, TX 78712-1083, USA\\       
}  
\date{Accepted 
      Received ; 
      in original form  }
                                                                                  
\pagerange{\pageref{firstpage}--\pageref{lastpage}}
\pubyear{}

\maketitle

\label{firstpage}

\begin{abstract}

    Mid-infrared photometry of R Coronae Borealis stars obtained from various
 satellites from {\it IRAS} to {\it WISE} has been utilized in
  studying the variations of the circumstellar dust's contributions to
 the spectral energy distribution of
  these stars. The variation of the fractional coverage  (R) of dust clouds
  and their  blackbody  temperatures (T$_d$) have been used in trying to understand
   the  dust  cloud evolution over the three decades spanned by the satellite observations.
   In particular,  it is shown that a prediction R$ \propto T_d^4$  developed in the paper is satisfied, especially by those stars for
   which a single collection of clouds dominates the IR fluxes.

\end{abstract}


\section{Introduction}

R Coronae Borealis stars (here, RCBs) are a rare class of 
peculiar variable stars. Currently known stars number about 68 in the
Galaxy, 19 in Large Magellanic Cloud and three in the Small Magellanic Cloud  (Clayton 2012).
Two principal defining
characteristics of RCBs are (i) a propensity to fade at unpredictable times
by up to about
8 magnitudes as a result of  clouds of soot intercepting the line of 
sight to the star, and
(ii) a supergiant-like atmosphere that is
very H-deficient and He and C-rich.

RCB stars manufacture and disperse soot from a formation site
near the star (Loreta 1934, O'Keefe 1939).  The soot particles absorb star light and reemit the energy
in the infrared (e.g., Feast et al. 1997) . Often, soot particles are sufficiently
 close to the
star that they achieve an equilibrium temperature of several hundred
degrees and so emit in the mid-infrared bands. Such emission is measured
with difficulty from the ground. Fortunately, several Earth-orbiting
satellites built for infrared photometric surveys have now reported
measurements on RCBs. Available data are too sparse and inhomogeneous to
define the variation of RCB dust emission on a daily, monthly or even
annual timescale but mean properties of the infrared emission should
be extractable to enable searches for correlations between dust shell
properties derived from the infrared emission and other RCB properties
such as chemical compositions.

The {\it Infrared Astronomical Satellite, IRAS} provided the 
first systematic photometry of these
stars (Walker 1986; Rao \& Nandy 1986) to complement 
limited ground-based
  photometry carried out earlier (e.g., Feast \& Glass 1973; 
Kilkenny \& Whittet
  1984). {\it IRAS} confirmed not only that warm (300--900K) dust was
common around RCBs but also showed the presence of cold (i.e., distant from 
the star) dust for strong infrared emitting stars (Rao \& Nandy 1986). 
{\it IRAS} was followed by the {\it Infrared Space Observatory, ISO}.
Although {\it ISO} provided higher spectral resolution mid-infrared
  spectroscopy and SEDs, it observed only a few bright RCBs (Lambert et al.
  2001).  Clayton et al. (2011) recently discovered using the {\it Herschel} satellite
  that the cold dust shell  around  R CrB  radiates up to at least a wavelength of 1000 $\mu$m.

              Major progress   in infrared photometry of RCBs could  occur only  
after the launch of
{\it  Spitzer} satellite which provided high-quality low-resolution 8-40 $\mu$m
spectra using IRS for a major
  sample of RCBs (Garc\'{i}a-Hern\'{a}ndez et al. 2011, here Paper I). 
{\it Spitzer} spectra were combined with optical and near-infrared photometry at
maximum light corrected for interstellar reddening to define a star's
spectral energy distribution (SED) from which characteristics of the infrared
emitting dust shell were extracted (Garc\'{\i}a-Hern\'{a}ndez et al. 2011, 2013).
Recently, it was  realised that mid-infrared band colours
  could distinguish RCB stars from other types of variable   stars rather uniquely
  (Tisserand 2012; Miller et al. 2013; Tisserand et al. 2013).

Following {\it Spitzer}, the satellites {\it AKARI} and {\it WISE} have provided
photometric surveys from which observations of RCBs may be extracted.
The {\it AKARI}  infrared camera surveyed the sky at 9 and 18$\mu$m
between 2006 May 6 and 2007 August 28
(Ishihara et al. 2010). A few RCBs were also detected by {\it AKARI}
at 65, 90 and 160 $\mu$m. 
{\it WISE}'s survey was conducted at 3.0, 4.6, 12 and 22 $\mu$m
in 2010
(Wright et al. 2010). Tisserand (2012) used the available {\it AKARI},
{\it WISE}  and other photometry to investigate the infrared colours
and the spectral energy distributions from the optical to the mid-infrared
and to characterize the circumstellar dust emission from a large sample of
RCBs. 

Here, we assemble and 
discuss the SEDs  of our {\it Spitzer} sample and their dust
emission as measured in the mid-infrared by the series of satellites
from {\it IRAS} to {\it AKARI} and {\it WISE} and by published ground-based
photometry.

\begin{table*}
\centering
\begin{minipage}{180mm}
\caption{\large IRAS, AKARI, Spitzer, WISE flux densities (Jy) at selected wavelengths}
\begin{tabular}{lllrrrrrrrlllr}
\hline
\multicolumn{1}{c}{Star}&\multicolumn{1}{c}{}&\multicolumn{1}{c}{ }&\multicolumn{1}{c}{}&
\multicolumn{1}{c}{} &\multicolumn{1}{c}{} &\multicolumn{1}{c}{} &\multicolumn{1}{c}{}&\multicolumn{1}{c}{}&\multicolumn{1}{c}{}&\multicolumn{1}{c}{}&\multicolumn{1}{c}{}&\multicolumn{1}{c}{}&\multicolumn{1}{c}{}\\
\cline{2-14}  \\
& {\it IRAS$^a$} & &  & {\it AKARI}  &  &
 &Spitzer &  &  & & {\it WISE}    & \\
           & 1983 &  &  & 2006 &  &  &  &      &   & & 2010.5 & &        \\
\cline{2-3}  \cline{5-6}   \cline{8-10}  \cline{12-13}  \\
           &12$\mu$m & 25$\mu$m &   &9$^b$$\mu$m &18$\mu$m&  &9$\mu$m& 12$\mu$m&18$\mu$m &   &12$\mu$m&22$\mu$m &   \\
           &     &   &    &  &   &  &   &     & &   & &   &        \\
\hline
{\bf XX Cam}&0.23  &0.14 &  &     &1.39$\pm$0.06&  &       &     &     &  &1.21$\pm$0.02&0.61$\pm$0.01&  \\
{\bf SU Tau}&9.50$\pm$0.76& 4.14$\pm$0.29&  &14.72$\pm$0.50&6.16$\pm$0.05& &       &7.52$\pm$0.08&4.88$\pm$0.06&  &7.12$\pm$0.07&3.35$\pm$0.04&  \\
{\bf UX Ant} &     &     &  &0.10$\pm$0.01&     &  &       &     &     &  &0.08$\pm$0.00&0.04$\pm$0.00&  \\
{\bf UW Cen}&7.81$\pm$0.47&5.57$\pm$0.33&  &9.76$\pm$0.07&5.70$\pm$0.05&  & 9.78 &7.66$\pm$0.07&5.29&  &7.36$\pm$0.05&4.67$\pm$0.05&  \\
{\bf Y Mus}  &0.83$\pm$0.07& 0.29$\pm$0.03&  &0.23$\pm$0.02&     &  & 0.18 &0.18$\pm$0.01&0.14$\pm$0.01&  &0.13$\pm$0.00 &0.12$\pm$0.00 &  \\
{\bf DY Cen}& 1.05$\pm$0.06&0.84$\pm$0.07&  &0.54$\pm$0.02&0.89$\pm$0.02&  & 0.50 &0.87$^c$&1.07&  &0.66$\pm$0.01 &0.91$\pm$0.01&  \\
{\bf V854 Cen}&18.42$\pm$0.9&5.88$\pm$0.4&  &22.97$\pm$1.2&7.36$\pm$0.03& &16.36 &11.59$\pm$0.03&6.79$\pm$0.0& &17.78$\pm$0.2&5.83$\pm$0.04&  \\
{\bf Z UMi} & 1.79$\pm$0.11&0.62$\pm$0.04&  &1.72$\pm$0.03&0.76$\pm$0.02&  & 1.54 &1.18$\pm$0.02&0.69$\pm$0.05&  &1.23$\pm$0.01&0.56$\pm$0.01&  \\
{\bf S Aps} & 2.71$\pm$0.11&1.02$\pm$0.07&  &1.97$\pm$0.02&1.00$\pm$0.02&  & 3.14 &2.30$\pm$0.02&1.31$\pm$0.04&  &0.91$\pm$0.01&0.60$\pm$0.01&  \\
{\bf R CrB} & 33.83$\pm$1.4&13.81$\pm$0.6&  &52.99$\pm$2.4&21.48$\pm$0.03&  &       &23.72$\pm$0.3&13.66$\pm$0.4&  &45.95$\pm$0.8&17.52$\pm$0.1&  \\
{\bf RT Nor}&0.85$\pm$0.09&0.39$\pm$0.04&  &0.26$\pm$0.04&0.23$\pm$0.02&  & 0.15 &0.18$\pm$0.00&0.20$\pm$0.00&  &2.01$\pm$0.02&0.77$\pm$0.01&  \\
{\bf RZ Nor}&3.08$\pm$0.09&1.77$\pm$0.09&  &3.12$\pm$0.26&1.70$\pm$0.08&  & 3.63&2.93$\pm$0.01&2.01$\pm$0.01&  &2.05$\pm$0.03 &1.26$\pm$0.02&  \\
{\bf V517 Oph}&6.0$\pm$0.36&1.90$\pm$0.19&  &6.76$\pm$0.24&     &  &       &4.86$\pm$0.06&2.75$\pm$0.06&  &4.47$\pm$0.04&1.86$\pm$0.02 &  \\
{\bf V2552 Oph}&   &     &  &     &     &  &       &     &     &  &0.36$\pm$0.00&0.16$\pm$0.00 &  \\
{\bf V532  Oph}& 0.52$\pm$0.04&   &  &     &0.47$\pm$0.06&  &       &     &     &  &0.60$\pm$0.01 &0.30$\pm$0.01&  \\
{\bf V1783 Sgr}&2.84$\pm$0.14&1.00$\pm$0.09&  &3.92$\pm$0.25&1.90$\pm$0.03&  & 2.97&2.50$\pm$0.02&1.66$\pm$0.04&  &1.62$\pm$0.02&0.79$\pm$0.02&  \\
{\bf WX CrA}&1.89$\pm$0.09&0.61$\pm$0.05&  &1.74$\pm$0.10&0.81$\pm$0.00&  & 0.89&0.73$\pm$0.02&0.53$\pm$0.03&  &0.90$\pm$0.01&0.41$\pm$0.01&  \\
{\bf V739 Sgr}&1.10$\pm$0.09&0.28$\pm$0.04&  &1.64$\pm$0.06&0.84$\pm$0.00&  & 1.45 &1.20$\pm$0.01&0.79$\pm$0.02&  &0.86$\pm$0.01&0.42$\pm$0.01&  \\
{\bf V3795Sgr}&3.42$\pm$0.27& 1.36$\pm$0.20&  &4.57$\pm$0.14&1.95$\pm$0.07&  & 3.31 &2.63$\pm$0.05&1.57$\pm$0.04&  &1.70$\pm$0.02&0.93$\pm$0.01&  \\
{\bf VZ Sgr}&1.13$\pm$0.08& 0.60$\pm$&  &     &0.28$\pm$0.10&  & 1.30&1.08$\pm$0.02&0.86$\pm$0.04&  &0.30$\pm$0.00&0.16$\pm$0.00&  \\
{\bf RS Tel}&1.31$\pm$0.08&0.57$\pm$0.05&  &1.87$\pm$0.10&0.77$\pm$0.01&  & 1.82 &1.33$\pm$0.00&0.90$\pm$0.00&  &0.96$\pm$0.01&0.51$\pm$0.01&  \\
{\bf MACHO$^d$}& & & &    &     &  & 0.54 &0.45&0.32&  &0.27$\pm$0.00 &0.17$\pm$0.00&  \\
{\bf GU Sgr}&0.97$\pm$0.07&0.66$\pm$0.14&  &1.29$\pm$0.14&0.75$\pm$0.1&  &       &     &     &  &1.74$\pm$0.02&0.76$\pm$0.01&  \\
{\bf NSV11154}&0.41$\pm$0.02&0.17$\pm$0.02&  &0.63$\pm$0.01&0.30$\pm$0.03&  &       &     &     &  &0.24$\pm$0.00 &0.15$\pm$0.00&  \\
{\bf MV Sgr}& 0.86$\pm$0.19&1.48$\pm$0.13 &  &0.33$\pm$0.02&1.01$\pm$0.01&  & 0.32 &0.56$\pm$0.01&1.06$\pm$0.03&  &0.38$\pm$0.01 &1.06$\pm$0.02 &  \\
{\bf FH Sct} &0.54$\pm$0.05& 0.40$\pm$0.05&  &1.40$\pm$0.03&0.66$\pm$0.05&  & 1.02 &0.92$\pm$0.01&0.68$\pm$0.02&  &1.06$\pm$0.01&0.52$\pm$0.01 &  \\
{\bf V CrA} &4.95$\pm$0.20&2.00$\pm$0.14&  &3.61$\pm$0.21&2.17$\pm$0.01&  & 3.32 &3.02$\pm$0.00&2.40$\pm$0.01&  &3.46$\pm$0.03&1.96$\pm$0.02 &  \\
{\bf SV Sge}& 3.29$\pm$0.23&1.66$\pm$0.10&  &3.85$\pm$0.44&2.29$\pm$0.10&  & 2.15 &2.10$\pm$0.02&1.69$\pm$0.03&  &1.36$\pm$0.01&1.10$\pm$0.02&  \\
{\bf V1157 Sgr}&2.63$\pm$0.21&0.89$\pm$0.08&  &2.57$\pm$0.00&1.14$\pm$0.02&  & 2.29 &1.74$\pm$0.02&1.08$\pm$0.03&  &1.64$\pm$ &0.79$\pm$ &  \\
{\bf RY Sgr}& 63.88$\pm$4.5&20.80$\pm$0.8&  &48.00$\pm$3.7&20.18$\pm$1.0&  &       &26.77$\pm$0.4&17.25$\pm$0.3& &33.16$\pm$0.9&13.89$\pm$0.2&  \\
{\bf ES Aql}&1.18$\pm$0.09&0.39$\pm$0.07&  &1.94$\pm$0.13&0.88$\pm$0.04&  & 1.77 &1.34$\pm$0.01&0.87$\pm$0.02&  &1.34$\pm$0.02 &0.61$\pm$0.01&  \\
{\bf V482 Cyg}&0.85$\pm$0.05& 0.35$\pm$0.07&  &1.13$\pm$0.04&0.46$\pm$0.02&  & 0.58 &0.49$\pm$0.00&0.38$\pm$0.00&  &0.72$\pm$0.01&0.33$\pm$0.01&  \\
{\bf U Aqr} &1.12$\pm$0.09&     &  &1.26$\pm$0.14&0.75$\pm$0.00&  & 0.89&0.86$\pm$0.01&0.82$\pm$0.02&  &0.52$\pm$0.01&0.37$\pm$0.01&  \\
{\bf UV Cas} &3.81$\pm$0.15& 1.28$\pm$0.09&  &1.13$\pm$0.05&0.80$\pm$0.03&  & 0.87 &0.83$\pm$0.01&0.65$\pm$0.02&  &0.60$\pm$0.01&0.49$\pm$0.01&  \\
\hline
\end{tabular}
\\
$^{a}$: Color corrected. Mainly from Walker (1986). \\
$^{b}$: Uncertainties are less than 0.01 Jy for all stars. 
$^{c}$: Emission features exist at 12$\mu$m and 18$\mu$m in the {\it  Spitzer} spectrum of DY Cen (Garc\'{\i}a-Hern\'{a}ndez et al. 2011, 2013).
$^{d}$: MACHO135.27132.51 \\

\end{minipage}
\end{table*}

 \section{Spectral Energy Distributions}

In Paper I, {\it Spitzer} satellite spectra of  
 RCBs in the wavelength range of 5 to 40 $\mu$m were paired
 with visual to near-infrared photometry of the star to obtain a comprehensive SED of the star.  Observed fluxes
 were corrected for interstellar reddening. For details, please see Paper I.
Comparisons were made with {\it IRAS} 10 and 25 $\mu$m fluxes and, where available with ground-based infrared photometry.
  Here, we complement
 these {\it Spitzer}  and {\it IRAS}
 observations with  {\it AKARI} (Ishihara et al. 2010) and 
{\it WISE} (Wright et al. 2010)
 satellite band fluxes (Table 1). 
 {\it WISE} observations were done in four bands 3.0, 4.6, 12 and 22 microns 
whereas
{\it AKARI} observations in the two bands at 9.0 and 18.0 microns have been
 used. The flux calibration adopted was from Wright et al. (2010) for {\it WISE}
 magnitudes and Ishihara et al. (2010) for {\it AKARI}. The epochs of these 
 observations have been taken as 2006.0 and 2010.5 for {\it AKARI} and
 {\it WISE}, respectively. Epochs of the {\it Spitzer} observations are given in Paper I: a majority were taken in about mid-2008.
 To compare the flux densities of {\it Spitzer} spectra with the other satellites
   bands we choose the the same effective wavelengths 
 of these
 bands 9, 12 , 18 $\mu$m and obtained monochromatic 
 flux densities from the {\it IRS} spectra . We made a linear fit to the data points
  of three wavelengths on either side of these chosen band wavelengths and
  obtained the flux density listed in Table 1 at these central wavelengths. These flux densities
  are not band averaged flux densities as given by other satellites.
  Flux density uncertainties were taken from the following sources: {\it IRAS} Point Source Catalogue 2.0  via SIMBAD;
 {\it AKARI} and {\it WISE} from http://irsa.ipac.caltech.edu/cbi-bin/Gator/nph-query; {\it Spitzer} from the noise in the observed
 spectra.  For the {\it Spitzer} 9 $\mu$m flux density, the uncertainties are less than 0.01 Jy for all stars and are not indicated in
 Table 1.    {\it WISE} measurements  at 3.0 and 4.6 microns are
 not tabulated but are shown in the appropriate figures which follow.  Except for the {\it IRAS} photometry,  the ground-based and {\it AKARI}
 and {\it WISE} photometry have not been colour  corrected.  Apart from the 12 $\mu$m  {\it WISE} band, the other bands are sufficiently
 narrow  not to require a significant colour correction for these spectra which span a narrow (500 -- 900 K) temperature range.

  Earlier, Tisserand (2012) 
 used {\it WISE}  as well as {\it AKARI} magnitudes to construct SEDs of  RCBs.
The recent release of the {\it WISE} catalogue contains more and fainter RCBs
 than dealt  with by him and the magnitudes also were revised (IPAC website 
 http://irsa.ipac.caltech.edu/cbi-bin/Gator/nph-query). 
It was mentioned
 by  Tisserand (2012) that 4.6 micron band {\it WISE}
 magnitudes less that 4.0 have 
 saturation problems. Some of the bright RCBs
 like R CrB, RY Sgr etc. were too bright for accurate photometry. 
 Table 1 gives the flux density measurements  from the four satellites and, where available, ground-based mid-IR photometry.
  We make limited use of the {\it WISE} flux densities at 3.0 and 4.6 $\mu$m. These quantities are not tabulated in Table 1 but are provided by 
  Tisserand (2012).

 We have used the methodology described in Paper I
 for constructing the SED of the star and its dusty circumstellar shell. Interstellar reddening estimates  listed in Paper I 
 have been applied  before constructing the SEDs; this correction for reddening primarily affects the stellar component of the
 combined SED. Blackbody fits have been made to the combined SEDs with the temperature of the stellar blackbody taken from
 Paper I. 
                 
\begin{table*}
\centering
\begin{minipage}{180mm}
\caption{\large Dust Parameters from MidIR SEDs of RCBs at various epochs.}
\begin{tabular}{llllllllllllll}
\hline
{}     &     \multicolumn{3}{c}{Ground/Other}&\multicolumn{2}{c}{\it IRAS}&\multicolumn{2}{c}{\it AKARI}& \multicolumn{3}{c}{Spitzer}&
\multicolumn{2}{c}{\it WISE}& \multicolumn{1}{c}{Mean R}\\
{Star}     &     \multicolumn{3}{c}{}&\multicolumn{2}{c}{1983}&\multicolumn{2}{c}{2006}& \multicolumn{3}{c}{}&
\multicolumn{2}{c}{2010.5}& \multicolumn{1}{c}{}\\
\cmidrule(rl){2-4}  \cmidrule(rl){5-6}  \cmidrule(rl){7-8}  \cmidrule(rl){9-11}  \cmidrule(rl){12-13}\\
           &Epoch&R&T$_d$  &R&T$_d$  &R&T$_d$  & Epoch&R&T$_d$   &  R&T$_d$ & R$_{\rm av}$  \\
           &     &   &    &  &   &  &   &     & &   & &   &        \\
\hline
{\bf XX Cam}&    &   &    &$\ll$0.01&500&0.02&550&      &   &   &0.01&650& 0.01$\pm$0.01  \\
{\bf SU Tau}&    &   &    &0.46&600&0.99&850& 2008.4&0.44&635&0.54&750&0.61$\pm$0.23 \\
{\bf UX Ant}&    &   &    &     &   &0.14&650&        &    &   &0.14&650&0.14          \\
{\bf UW Cen}&1984$^{e}$&0.35&620&0.44&630&0.44&600&2008.7&0.44&636&0.44&650&0.38$\pm$0.10 \\
{\bf Y Mus} &1984$^{b}$&0.14&980&0.08&655&0.01&420&  2008.4&0.01&395&0.01&340&0.06$\pm$0.06 \\
{\bf V854 Cen}&1996.8$^{a}$&0.94&1100&0.57&920& 0.54&900&2007.8&0.30&900&0.54&880&0.58$\pm$0.22 \\
{\bf Z UMi}&     &   &    &0.95&850 & 0.79&900 &2008.9&0.41&695&0.79&900&0.74$\pm$0.20 \\
{\bf S Aps}&     &   &    &0.42 & 750 & 0.22& 710&2008.4&0.37 &730&0.14&950& 0.29$\pm$0.11 \\
{\bf R CrB}&1998.0$^{g}$&0.30&810&0.20&680&0.30&810&2004.6&0.30&950&0.30&810&0.28$\pm$0.04\\
{\bf RT Nor}&1984$^{b}$&0.11&920&0.08&500&0.02&400&2005.4&0.01&365&0.46&850&0.14$\pm$0.17 \\
{\bf RZ Nor}&1983.4$^{c}$&0.72&720&0.72&720&0.54&650&2006.3&0.64&640&0.37&610&0.57$\pm$0.13 \\
{\bf V517 Oph}&   &   &    &0.98& 850 & 0.92&885  &2008.4&0.92&885&0.83&880&0.91$\pm$0.05 \\
{\bf V2552 Oph}& &   &    &     &   &     &   &      &   &   &0.09&800& 0.09   \\
{\bf V532  Oph}$^{f}$& &  &     &0.03&400&     &   &      &   &   &0.07&720& 0.05$\pm$0.02 \\
{\bf V1783 Sgr}&   &   &    &0.53&770 &0.65&790  &2008.4&0.27&554&0.24&670&0.42$\pm$0.17 \\
{\bf WX CrA}&    &   &    &0.46& 670 & 0.39&740  &2008.9&0.14&570&0.35&870&0.33$\pm$0.12 \\
{\bf V739 Sgr}&  &   &    &0.98& 900 & 0.71&700  &2008.4&0.57&656&0.68&850&0.74$\pm$0.15 \\
{\bf V3795Sgr}&  &   &    &0.50&700&0.36&650& 2008.4&0.29&600&0.17&580&0.33$\pm$0.12 \\
{\bf VZ Sgr}&    &   &    &0.21&700&0.11&780 & 2008.4&0.21&692&0.07&770&0.15$\pm$0.06  \\
{\bf RS Tel}&1984$^{b}$&0.26&790&0.28&700&0.36&850&2005.78&0.33&830&0.15&650&0.28$\pm$0.07 \\
{\bf MACHO135.27132.51}& & & &   &     &            &      &     &   &     &   &               \\
{\bf GU Sgr}&1977.6$^{d}$&0.45&910&0.06&400&0.12&600&     &   &  &0.45&910&0.27$\pm$0.18   \\
{\bf NSV 11154}&  &   &    &0.63& 720 & 0.92&800  &      &     &   &0.40&750&0.65$\pm$0.21 \\
{\bf FH Sct} &   &   &    &0.04&393& 0.17&740&  2008.5&0.10&537&0.25&940 & 0.14$\pm$0.09 \\
{\bf V CrA}&1984$^{b}$&0.69&640&0.87&670&0.45&580&2005.4&0.38&550&0.65&700&0.61$\pm$0.18 \\
{\bf SV Sge}&    &   &    &0.17&730 &0.11&600  &2008.5&0.05&500&0.03&500&0.09$\pm$0.05 \\
{\bf V1157 Sgr}& &   &    &0.89& 880 & 0.60&820  &2008.5&0.44&753&0.34&680&0.57$\pm$0.21 \\
{\bf RY Sgr}&1997.3$^{h}$&0.38&820&0.76&870&0.38&820&2004.89&0.20&675&0.38&820&0.42$\pm$0.18 \\
{\bf ES Aql}&    &   &    &0.70& 830 & 0.70&830  &2008.5&0.58&750&0.70&830&0.67$\pm$0.06\\
{\bf V482 Cyg}&  &   &    &0.09&650&0.14&850& 2004.96&0.04&580&0.20&970&0.12$\pm$0.06 \\
{\bf U Aqr}&    &   &    &0.37&560  &0.37& 560& 2008.6&0.25&473  &0.24&680 &0.31$\pm$0.06 \\
{\bf UV Cas}&    &   &    &0.26&830&0.03&507& 2008.5&0.03&507&0.02&510&0.08$\pm$0.10 \\
\hline
\end{tabular}
$^{a}$ V854 Cen was observed by {\it ISO} on 1996 September (Lambert et al. 2001). The R and T$_d$ values refer to ISO observations. The {\it AKARI} and {\it WISE} photometry needed a
combination of two black bodies to match the SED.\\
$^{b}$ The ground based observations come from Kilkenny \& Whittet (1984). Their observations
were obtained mainly in April 1983 . \\
$^{c}$ The KW's and {\it IRAS} observation together defined the SED from which R and T$_d$ has been estimated. \\
$^{d}$ The ground based observations are from Glass (1978). \\
$^{e}$ Goldsmith et al. (1987) also observed the star on 1985.5 from which R and T$_d$ of 0.16 and  540K have been derived. They have been included in the average.\\
$^{f}$ Both {\it IRAS} 12$\mu$m and {\it AKARI} 18$\mu$m   fluxes together  defined a
 black body SED for the R and T$_d$ estimate.\\
$^{g}$ R CrB was observed by ISO on 1998 January 15. The {\it ISO} SED could be fit by a combination of blackbodies 810 K, 750K and 550K. The flux densities from {\it AKARI} and
{\it WISE}  match very well the ISO spectrum. All three observations have same value of R . \\
$^{g}$ RY Sgr was observed by ISO on 1997 March 25. The ISO spectrum could be fit with a single black body of 870K. The flux densities obtained with {\it AKARI} and {\it WISE}  match the {\it ISO} spectrum very well. All three observations result in same R. \\
\end{minipage}
\end{table*}

The blackbody fits provide the dust temperature T$_d$ and the fraction of the
stellar energy radiated by the cool circumstellar shell R$ = f_{cool}/f_{star}$  where $f$ refers to the integrated flux density.  In general, 
 the dust component for RCBs can be well represented by a single
blackbody. In a few cases. a second and cooler blackbody is added to provide an adequate fit to the IR fluxes. With a more
detailed coverage of the near-IR energy distribution, it might be possible
to include a blackbody to represent emission by warmer dust near the
star.  The {\it WISE} observations at 3.0 and 4.6 $\mu$m and ground-based L and M photometry sample warmer
dust, if present. Our focus
here is on the dust emitting at mid-IR wavelengths.

\section{Discussion - the puff model}

\subsection{The concept}

Our discussion of SEDs is framed in terms of the random dust-puff model for
dust formation and ejection. In this scenario (Feast 1979, 1986, 1996, 1997;
Feast et al. 1997; Herbig 1949; Payne-Gaposchkin 1963; Clayton et al. 1992), dust is
ejected from points on or near the RCB surface in the form of puffs or clouds
which then are expelled  away from the stellar surface. 
Only those puffs formed on or very close to the Earth-facing hemisphere of
the RCB will result in the characteristic fading. All puffs but those eclipsed by the star will
contribute to the infrared signal from the dust.
In a given circumstellar
envelope, there may be one, several or many puffs. Puffs may be
created at preferred parts of the stellar surface or uniformly over the
surface. The rate of creation of fresh puffs will likely vary from
star to star. Those RCBs prone to frequent formation of puffs will
be expected to have nearly time-independent $R$ values.  The amplitude of IR variations will depend 
on several factors such as the number of puffs and their mean contribution to the mid-IR flux: a star with many
weak puffs is likely to show smaller variations than a star with a few strong puffs. On the other hand,
RCBs producing puffs infrequently are likely to have smaller mean $R$ values
with larger variations resulting as a puff moves away from the star and subsequently  a fresh puff is ejected. 
In this simple picture,  puffs are formed and expand away from the star, i.e., there is no mechanism for storing dust (and gas)
near the star as in a circumstellar disk.  Chesneau et al. (2014)  from interferometric observations of V854 Cen show that this
star's dust is  concentrated in an elongated structure  which,  as the authors note, may reflect a bipolar wind off the star.  Quasi-stable dusty disks
have been found as circumbinary features for some binary stars. It has yet to be shown that V854 Cen is a binary star. Evidence for bipolar outflows have 
 been presented earlier by Rao \& Lambert (1993) for V854 Cen and Clayton et al. (1997) for R CrB.

Information about the formation  and evolution of the circumstellar dust
shell for RCBs is, thus, provided by the temporal evolution of the
infrared flux in terms of its spectral distribution and intensity
which here are crudely represented by the covering factor $R$ and the
blackbody temperature $T_d$.  Unfortunately,  available infrared measurements are extremely sparse because  the  observations are 
difficult to obtain; ground-based mid-IR photometry is a rarely
practiced art and infrared-capable satellites are launched infrequently.
An exceptional database is 
the  SAAO series of JHKL photometry of bright southern
RCBs (Feast et al. 1997).  And, in particular, Bogdanov et al. (2010) mounted a beautiful long-term campaign of JHKLM
observations of UV Cas.

\subsection{Variability}

Here, we compare mid-IR fluxes obtained with the satellites {\it IRAS}, {\it AKARI}, {\it Spitzer} and {\it WISE}. The epochs of these surveys
are  1983 for {\it IRAS},   2006 for {\it AKARI}, 2005-2008 for {\it Spitzer} and 2010 for {\it WISE}.  Long-term -- two decades approximately -- variations are estimated
from comparison between {\it IRAS} and the other three surveys. Shorter term -- one to a few years -- variations  are estimated by comparisons among the fluxes
from the three 
most recent satellites.
Our present  comparisons
are an extension of the {\it IRAS} -- {\it Spitzer} comparison  discussed in Paper I where 
 the {\it Spitzer} 12 and 25 $\mu$m
fluxes for a large RCB sample were compared with their {\it IRAS} counterparts obtained about
 25 years previously. 

{\it IRAS} and {\it Spitzer} 12$\mu$m flux densities are compared in Figure 1. When a few outliers are excluded,  the mean relation
$F(IRAS) = 1.24F(Spitzer)$ provides a good fit to the data points and this linear relation is shown in the figure.  A systematic offset  of 24 per cent between
{\it Spitzer} and {\it IRAS}  observations  may be due to the broad-band nature of the {\it IRAS}  photometry and an imperfect color correction.
Given that there is a roughly 10 per cent uncertainty affecting {\it IRAS} photometry, it would appear that many RCBs had similar 12 $\mu$m flux
densities in 1983 and 2005-2008 but this is not to suggest that they did not vary in the nearly 30 year interval.
  Outliers include RY Sgr, V854 Cen, UV Cas, Y Mus, RT Nor and WX CrA with an exceptionally large  ratio of  {\it IRAS} to {\it Spitzer} fluxes and FH Sct with  an apparently small ratio of {\it IRAS} to {\it Spitzer} fluxes.
The three stars -- UV Cas, Y Mus, and RT Nor -- showed
a factor of five greater {\it IRAS}  12 $\mu$m  flux than recorded by 
{\it Spitzer}.  A striking and expected
characteristic  of the trio is that 
their covering 
factors R as obtained from {\it Spitzer} were very small 
(0.01 to 0.03) and among
the lowest of the sample. For their {\it IRAS} fluxes, the R values were
0.07 -- 0.28.  In the puff model, the trio are most likely examples where very few puffs inhabit the circumstellar environment at any given time because
ejection of fresh puffs is an infrequent event.
Other outliers --RY Sgr, V854 Cen and WX CrA -- show a less extreme difference between {\it IRAS} and
{\it Spitzer} fluxes.  This trio have larger R covering factors but would appear to be still affected by a change in the dust in their
circumstellar shell. 
FH Sct , another low R star, is the only star in the sample which was fainter when observed by {\it IRAS} than by {\it Spitzer}.

A comparison of {\it Spitzer} and {\it AKARI} flux densities at 9 $\mu$m is shown in Figure 2 where the relation
$F({\it AKARI}) = F({\it Spitzer})$ is plotted.  Similarly,  {\it WISE} and {\it Spitzer} 12 $\mu$m fluxes are compared in
Figure 3 where the line corresponds to $F({\it WISE}) = F({\it Spitzer})$. In both figures, a few outliers are marked.   
In Figure 3, RT Nor is the outstanding outlier with its {\it WISE} flux density raised by a fresh ejection of dust. V854 Cen is again
brighter than expected from its {\it Spitzer} flux density and the mean for the sample.  Two stars -- VZ Sgr and S Aps -- have faded noticeably between their
observations by {\it Spitzer} and  then by {\it WISE}.   An overall impression from these comparisons is that the mid-IR fluxes of the majority of the sampled
RCBs are not subject to major variations over the 1 to 25 year timeframe. 

 \begin{figure}
\epsfxsize=8truecm
\epsffile{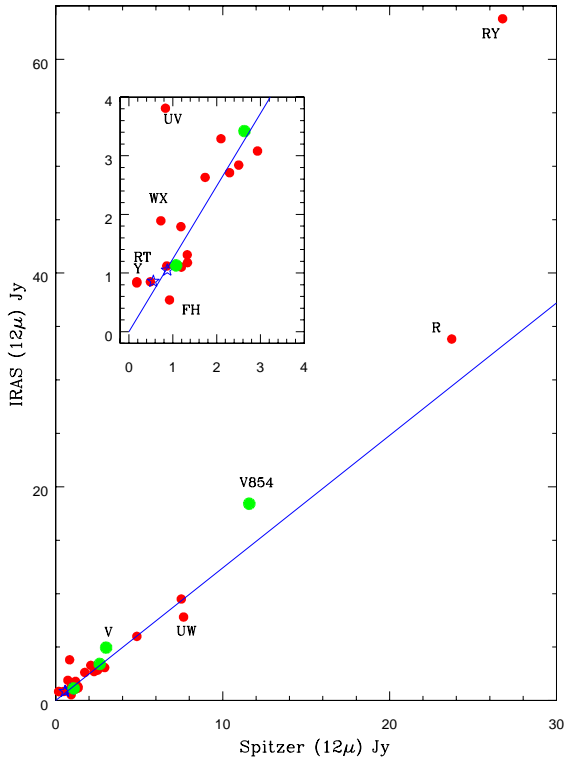}
\caption{ Flux densities  $F$ at 12$\mu$m from {\it Spitzer} versus the values from {\it IRAS}. The solid line shows the mean relation $F({\it IRAS}) = 1.24F({\it Spitzer})$
defined by the bulk of the sample.   Outliers are identified by shorthand labels: RY Sgr, V854 Cen, UV Cas, WX CrA, Y Mus, RT Nor and FH Sct. Red points refer to majority
RCBs as defined by Lambert \& Rao (1994), the green points refer to minority
RCBs and two blue stars refer to two hot RCBs  (DY Cen and MV Sgr) given in Table 1. }  
\end{figure}

 \begin{figure}
\epsfxsize=8truecm
\epsffile{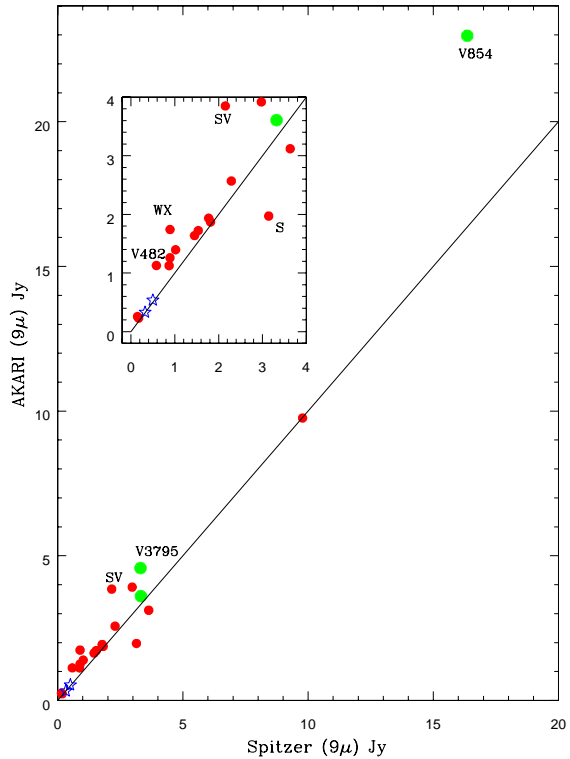}
\caption{ Flux densities  $F$ at 9$\mu$m from {\it Spitzer} versus the values from {\it AKARI}.  The  line shows the relation for $F({\it AKARI}) = F({\it Spitzer})$  Distinct outliers are identified by shorthand labels:  WX CrA, V482 Cyg , SV Sge, V854 Cen and S Aps. Note that there is no systematic shift between AKARI and
Spitzer flux densities. The symbols are same as in Figure 1. } 
\end{figure}

 \begin{figure}
\epsfxsize=8truecm
\epsffile{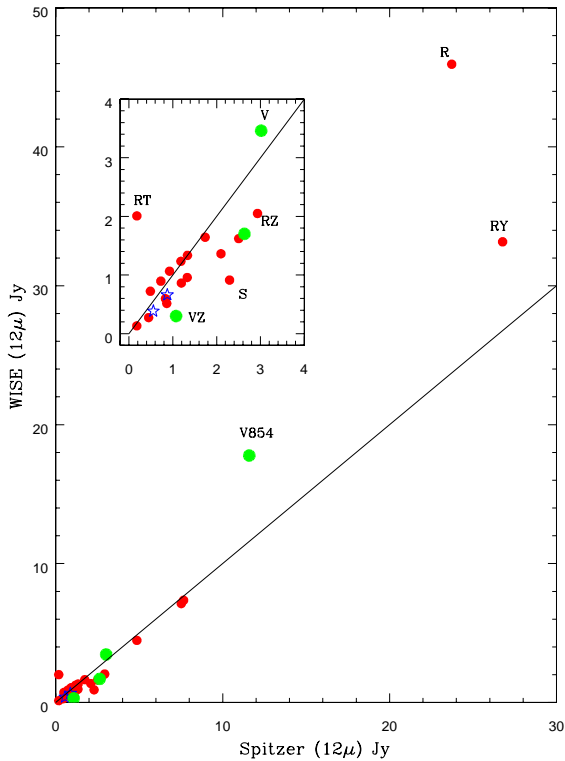}
\caption{ Flux densities  $F$ at 12$\mu$m from {\it Spitzer} versus the values from {\it WISE}. The solid line shows the mean relation $F({\it WISE}) = F({\it Spitzer})$
defined by the bulk of the sample.  Distinct outliers are identified by shorthand labels: notably RT Nor, VZ Sgr , R CrB and S Aps.  Symbols are as in Figure 1. } 
\end{figure}

Qualitatively, the three extreme outliers  in Figure 1 -- UV Cas, Y Mus and RT Nor -- 
with the
small R-values from {\it Spitzer} may be identified as stars which at the time
of the {\it Spitzer} observations had
little warm dust -- few dust puffs -- in their circumstellar envelopes and, thus, large
fractional changes in their
mid-IR fluxes  surely reflect
fresh ejection of dust into the envelope and possibly a single ejection on or off the line of sight. We begin comparison of the SEDs constructed from {\it IRAS},
{\it Spitzer} and {\it AKARI}, and {\it WISE} observations  with these three  {\it IRAS} -- {\it Spitzer} outliers -- see
Figure 4 for  UV Cas ,  Figure 5 for Y Mus and Figure 6 for RT Nor. 
 
For UV Cas, a valuable series of
{\it JHKLM} photometry by Bogdanov et al. (2010)
from 1984  (just following 
the {\it IRAS} observations)
to 2009  (just following the {\it Spitzer} observations) suggests that a major episode of dust ejection
occurred prior to the {\it IRAS} observations. This ejection did not cause
a fading of the star.  Following the ejection, UV Cas's mid-IR flux decayed
for about 2000 days with only two minor flux increases prior to 2009. Thus, the
dust detected by  {\it AKARI}, {\it Spitzer}  and {\it WISE} was likely from the pre-{\it IRAS}
ejection which had expanded away from the star in the intervening 25 years; 
the drop in dust temperature from 800 K to 510 K is consistent with this
idea (Paper I). 
The dates of the {\it AKARI} and {\it WISE} observations 
generally bracket the {\it
Spitzer} observations. Thus, it is not unexpected that for UV Cas
that mid-IR fluxes from the three satellites  are similar:  Figure 4 shows that
the {\it AKARI} and {\it Spitzer} fluxes are essentially identical  but the 
{\it WISE} fluxes are lower suggesting that the expansion of the 
`1980' dust ejection appears 
to have continued and may have accelerated.  A more quantitative discussion of the series of flux density measurements is given below 
in Section 3.3. However, Clayton et al. (2013) suggest that there might be
 some evidence for hundreds of km/s dust motions in the star.

\begin{figure}
\epsfxsize=8truecm
\epsffile{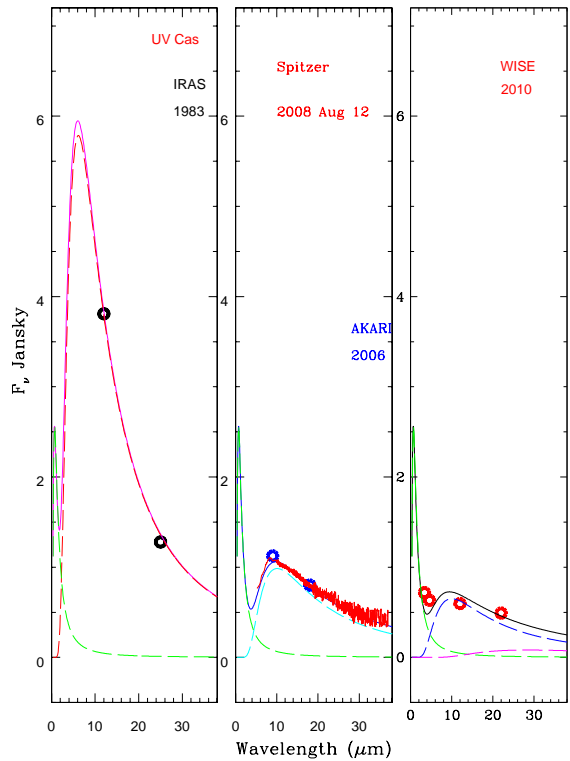}
\caption{ The SED for UV Cas  from three sets of infrared observations: left panel--{\it IRAS}; middle  panel- -{\it AKARI} and {\it Spitzer}; right panel - {\it WISE}. In each panel, a two or three black body fit is shown with a 7200K black body (green dashed curve) representing the stellar fluxes corrected for interstellar reddening  and  much cooler blackbodies representing the dust emission (830K for {\it IRAS},  507K  and 180K for {\it AKARI} and {\it Spitzer} and  510K and 180K for {\it WISE}). } 
\end{figure}

\begin{figure}
\epsfxsize=8truecm
\epsffile{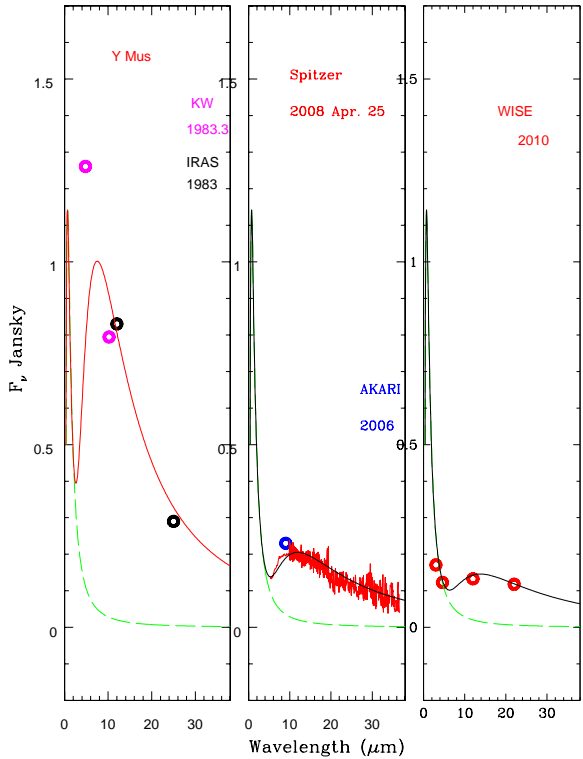}
\caption{The SED for Y Mus from IR observations at different epochs: left-panel - {\it IRAS} with M and N fluxes from Kilkenny \& Whittet (1984); middle
panel - {\it AKARI} and {\it Spitzer}; right panel - {\it WISE}. In each panel, a two  black body fit is shown  with a 7200K black body (green dashed curve)  representing the stellar
fluxes corrected for interstellar reddening and a cooler black body representing dust emission (655K for {\it IRAS},  420K for {\it AKARI} and 395K for
{\it Spitzer}, and 340K for {\it WISE}).
 }
\end{figure}

\begin{figure}
\epsfxsize=8truecm
\epsffile{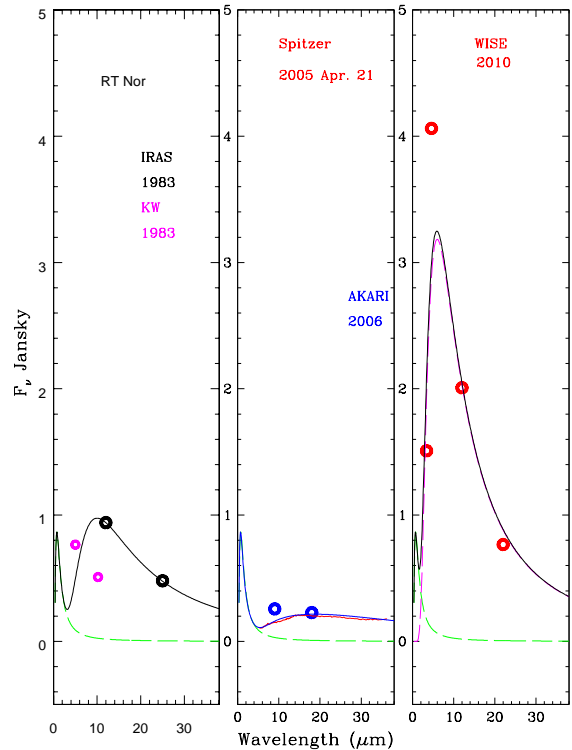}
\caption{The SED for RT Nor from IR observations at different epochs: :left-panel - {\it IRAS} with M and N fluxes from Kilkenny \& Whittet (1984);
middle panel - {\it AKARI} and {\it Spitzer}; right panel - {\it WISE}. In each panel, a two black body fit is shown with a 6700K black body representing
the stellar fluxes corrected for interstellar reddening and a cooler black body representing dust emission (500K for {\it IRAS},  400K for {\it AKARI}, 
and 850K for {\it WISE}).}
\end{figure}

{\it IRAS} to {\it WISE} photometry shows that Y Mus behaved similarly
to UV Cas (Figure 5): the {\it IRAS} fluxes and the N flux from Kilkenny \& Whittet (1984) are much greater than the fluxes from
{\it AKARI}, {\it Spitzer} and {\it WISE}. Kilkenny \& Whittet's M and N observations were made in 1983 April during the year-long
{\it IRAS} mission. Thus, the agreement between their N flux and the {\it IRAS} 10 $\mu$m flux is expected in the absence
of rapid changes of the IR flux (see discussion of RT Nor below). 
Unfortunately, there is no
long-term series of mid-IR photometry comparable to Bogdanov et al.'s (2010) 
observations of UV Cas but we presume that the dust detected by {\it IRAS}
moved away from Y Mus and had cooled substantially when detected
by the other satellites. {\it AAVSO} records show Y Mus has not experienced a
fading over the duration of the records (i.e., since early 1982) but these records do not sense puffs ejected off the
line of sight to the star.

In two respects,   different
circumstances are found for RT Nor (Figure 6). The {\it AKARI} and {\it Spitzer}
mid-IR fluxes of RT Nor
are in good agreement but the {\it WISE} fluxes are about an order
of magnitude stronger  and about twice the
{\it IRAS} fluxes.
The inference is that a fresh ejection of dust occurred after the 2005--2006
observations by {\it Spitzer} and {\it AKARI} and
prior to the observation in 2010 by {\it WISE}. 
This sharp increase in mid-IR fluxes may be linked to a fading of the star:
{\it AAVSO} records show that a
 minimum  occurred between October 2008 when the star was 
at maximum and the next observation in mid-2009 when it was about five
magnitudes below maximum, the first decline since 1990. 
A second aspect setting RT Nor apart from Y Mus is that Kilkenny \& Whittet's (1984) M flux is considerably
less than the {\it IRAS} measurement even though the M measurement was made during the year-long {\it IRAS}
mission. We conjecture that between the M and {\it IRAS} observations  fresh dust  was added to  the
circumstellar shell.  The AAVSO observations show no dimming in this period so that  dust replenishment must have occurred off
the line of sight to the star.   These few observations suggest that RT Nor is more active than either UV Cas or Y Mus.

Frequency of ejection of dust is likely related in some way to the
occurrence of visual fadings characteristic of RCBs but, as the examples of
UV Cas and Y Mus illustrate, not every ejection
of dust results in a dimming of the star. Jurcsik (1996) 
estimated inter-fade intervals for RCBs from reported visual magnitudes.
Her two longest intervals were for Y Mus (15300 days) and UV Cas (25500 days)
with typical values of 1000-2000 days reported for many RCBs. Thus,
the low R values, the large amplitude mid-IR flux variations, and the
long inter-fade intervals are all consistent with the view that UV Cas and
Y Mus are presently poor producers of dust. Jurscik's inter-fade interval for
RT Nor is listed as 1950 days. In Paper I, we noted that the last minimum
of RT Nor occurred about 15 years before 2008 and intimated that this suggested
that Jurcsik's 1950 day estimate  was an underestimate. However, recently
two minima have been observed with onsets in about January 2009 and February
2013. One may conjecture, as noted above,
 that the January 2009 fading was responsible for
the large increase 
in mid-IR flux between the {\it AKARI}$-${\it Spitzer}
 and {\it WISE}
observations.

\begin{figure}
\epsfxsize=8truecm
\epsffile{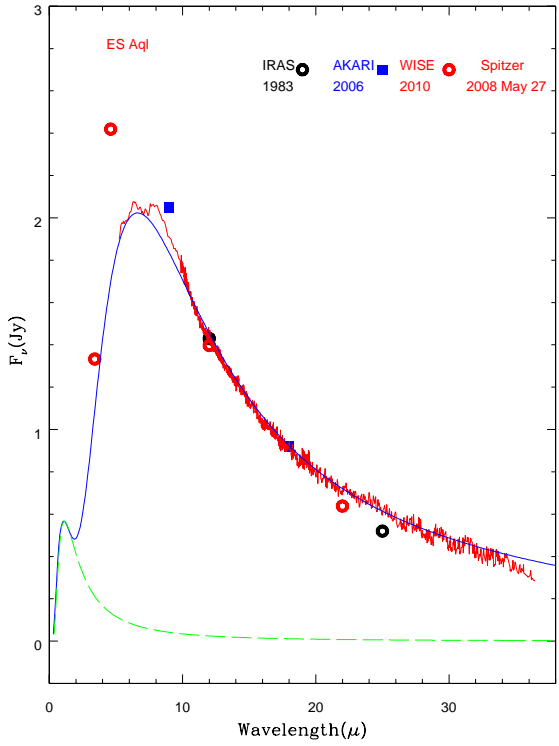}
\caption{SED of ES Aql. The fit  of two cool blackbodies (750K and 140K) is made to the {\it Spitzer} spectrum. {\it IRAS}, {\it AKARI}
and {\it WISE} observations show no sensible departures from the {\it Spitzer} spectrum except that the {\it WISE} flux density at 6$\mu$m 
exceeds the expectation from the {\it Spitzer} spectrum. The stellar blackbody corresponding to a temperature of 4500 K is shown by the green dashed curve.}
\end{figure}

At the opposite end of the range of  factors R to the above trio of outliers are  RCBs with large R. Among
this sample, there is a range of variation in the mid-IR fluxes from the four satellites.  At one extreme is ES Aql (Figure 7)
where the mid-IR fluxes are very similar from all four satellites.  The mean R is 0.67 $\pm$ 0.06.
  As noted in Paper I,  the cool RCB ES Aql declines frequently with a major decline
about every year. Unless these optical declines  reflect a preferred plane for puff ejection,  ES Aql's  circumstellar shell will be refreshed more
frequently than once a year. Thus, ES Aql is likely an example where the shell  hosts many puffs  and conditions in the shell are  essentially
time invariant. 
A similar case  may be the very active cool star
V517 Oph which shows an average R of 0.91 $\pm$ 0.05.

Other stars show modest variations in their mid-IR fluxes. One such example is SU Tau (Figure 8)
where the shape of the mid-IR spectrum is very similar from epoch to epoch but the flux level varies by about 
20-30\%. 
SU Tau is a star which is frequently in decline: it
`has been  three or more magnitudes below maximum light for nearly half of
the last 20 years' (Paper I). With frequent known ejections of dust and
quite possibly other ejections not causing fadings, it is not surprising that
the shape of the  mid-IR spectrum is almost invariant with modest changes in
strength. In the language of the puff model, SU Tau may resemble ES Aql but have fewer puffs in its circumstellar
shell  such that  a variation of one or two in the number of puffs emitting mid-IR radiation affects the total
emitted flux.

\begin{figure}
\epsfxsize=8truecm
\epsffile{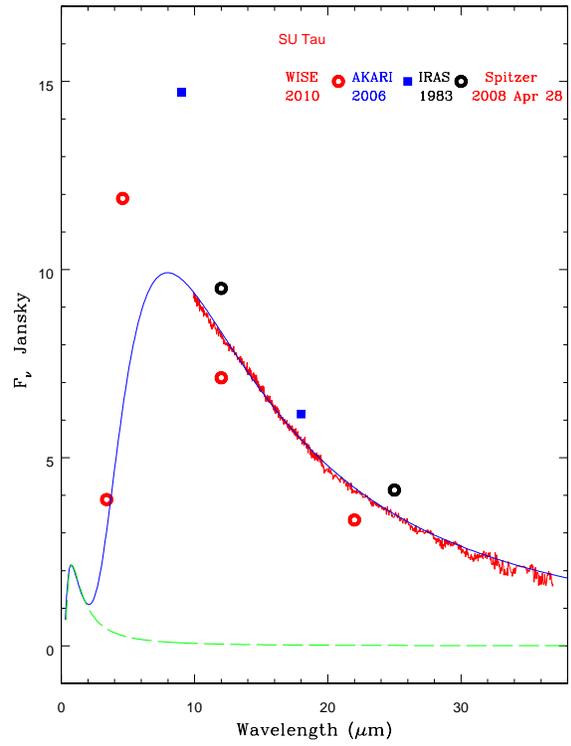}
\caption{SED of SU Tau. The fit  of a   blackbody at 635K is made to the {\it Spitzer} spectrum. {\it IRAS}, {\it AKARI}
and {\it WISE} observations show modest variations around the {\it Spitzer} spectrum with larger variations according to
{\it AKARI} and {\it WISE} at shorter wavelengths.  The stellar blackbody for 6500 K is shown by the green dashed curve.}
\end{figure}

 V3795 Sgr is a star for which {\it IRAS}, {\it Spitzer} and
{\it AKARI} give very similar mid-IR fluxes but {\it WISE} found appreciably
lower fluxes (Figure 9). V3795 Sgr, a minority RCB (Lambert \& Rao 1994),
has experienced just two declines in the last 20 years and during the period of
the {\it AKARI} - {\it WISE} observations was in recovery from one of those
declines.  A speculation is that the high speed wind seen during a recovery was responsible for sweeping a lot of dust outwards and, thus, to
cooler temperatures ( Rao et al. 1999; Clayton et al. 2013).
 
In the comparison with the {\it Spitzer} spectra obtained
between the times of the {\it AKARI} and {\it WISE} observations, 
 VZ Sgr (Figure 10), a minority RCB, stands out;
both the {\it AKARI} and the {\it WISE} fluxes are about
a factor of two to three less than the {\it Spitzer} and {\it IRAS}
fluxes. At the time of the {\it Spitzer} observation and as noted in
Paper I, VZ Sgr was in a deep minimum and several magnitudes below maximum
light, but, at the time of the {\it AKARI} and {\it WISE} observations, the
star was at maximum light. The high mid-IR flux seen by {\it Spitzer}
is likely attributable to the fresh dust contributed by the on-going minimum
but, in contrast to many other RCBs, the dust dispersed quickly; a high speed wind, as in V3795 Sgr, drove dust out
to great distances?.
  One may conjecture that in  cases  such as V3795 Sgr and VZ Sgr 
  puffs  form and/or
  dissipate on timescales of a couple of years or less. This behaviour  is in contrast to UV Cas, Y Mus and RT Nor where the timescale
  appears to be a couple of decades.

\begin{figure}
\epsfxsize=8truecm
\epsffile{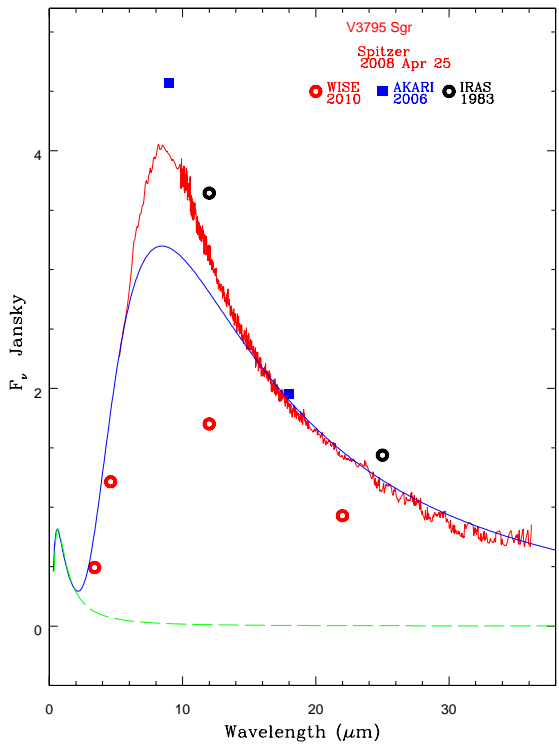}
\caption{SED of V3795 Sgr. The fit  of a blackbody at 600K is made to the {\it Spitzer} spectrum. {\it IRAS} and {\it AKARI} flux
densities are similar to the {\it Spitzer} spectrum but
the {\it WISE} observations show considerably lower fluxes with little change in the fitted blackbody temperature (580K versus 600K). The stellar blackbody of
8000 K is shown by the green dashed curve.}
\end{figure}

\begin{figure}
\epsfxsize=8truecm
\epsffile{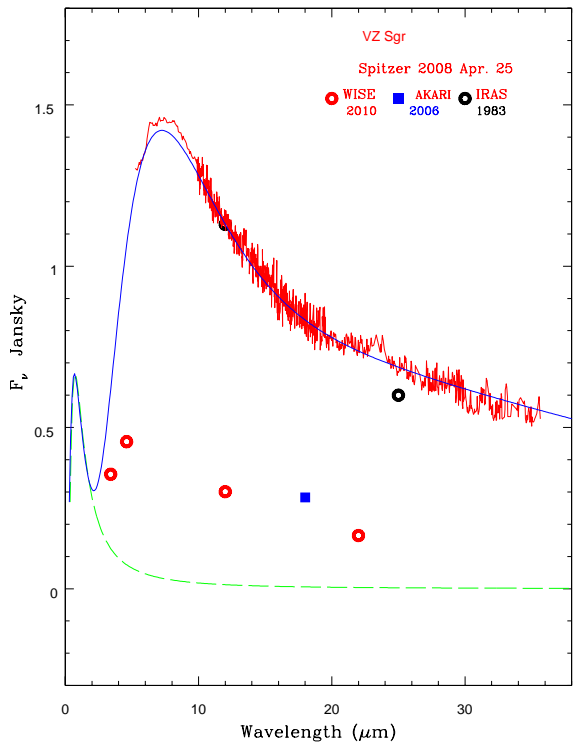}
\caption{SED of VZ Sgr. The fit  of a 690K blackbody is made to the {\it Spitzer} spectrum. {\it IRAS} flux densities fits this spectrum very well
but the {\it AKARI} and {\it WISE} measurements are factors of three to four lower.  The stellar blackbody of 7000 K is shown by the green dashed curve.} 
\end{figure}

\subsection{The run of R with $T_d$}

Simple implementations of the puff model suggest a correlation between the
covering factor R and the blackbody dust temperature T$_d$. 
The factor R is defined to be the ratio of integrated dust emission to the
stellar flux: R = $f_{dust}/f_{star}$ where corrections are applied for
interstellar reddening.
Consider
the following scenario. Suppose that puffs occupy a volume $V(r)$ at a
distance $r$ from the central star and the dust in the puffs is at an
equilibrium temperature T$_d$. The thickness of a puff is assumed to
be small relative to the radial distance $r$. The integrated emission from the
dust will be proportional to the product of $V(r)$ and T$_d^4$. The
dust temperature T$_d$ is taken to be the equilibrium temperature of a
grey dust grain in an optically thin puff\footnote{ The ratio of absorption coefficients for carbonaceous grains at 5000\AA\ to 20 $\mu$m is about a
factor of 50 (Colangeli et al. 1995) and, thus, all but the clouds causing the very deepest of declines will be optically thin
at mid-IR wavelengths.},i.e., T$_d \propto r^{-0.5}$
(Kwok 2007, p.314, Equation 10.32). In this simple picture, the ratio of
star's integrated dust emission from a given set of puffs moving out from 
distance $r_1$ to $r_2$ is $f(r_1)/f(r_2) =
V(r_1)/V(r_2) \times $(T$_d(r_1)$/T$_d(r_2))^4$. Thus, if the puffs 
are optically thin
at IR wavelengths  and their dust content are not evolving, $f_{dust}$ is expected
to scale with T$_d^4$.  Note that T$_d$ is determined from the 
wavelength-dependence of the IR emission and $f_{dust}$ from the integrated IR
emission and, thus, T$_d$ and $f_{dust}$ are different measures of the IR
emission.  

For the sample of RCBs, one expects that for a given T$_d$, there will
be a spread in $f_{dust}$ because the size and number of puffs and their distribution
with radial distance $r$  will vary from
star to star and time to time for a particular star. 
The temperature T$_d$ distance $r$ relation will depend also
on the stellar temperature T$_\star$: T$_d \propto$ T$_\star$ 
-- see Kwok above.   
Unfortunately, there is no observational way to determine the mean
radial distance of the puffs at a given time or the distance 
travelled by puffs between two observations without assuming any expansion
 velocity for the dust.  
The covering factor $R = f_{dust}/f_{star}$ involves the stellar flux
which is set by the luminosity $L$ or the product $R_{star}^2T_{star}^4$.
  
 Observational pursuit of the above idea may be most instructive when
applied to those stars in Table 2 for which R and T$_d$ both decrease
with time. In such cases, the reasonable inference is that a set of
puffs is expanding away from their formation site near the star and,
thus, experiencing cooling as observations proceeded
from {\it IRAS} to {\it WISE}. Six stars (Y Mus, UV Cas,
V3795 Sgr, SV Sge, V1157 Sgr and RT Nor) meet our condition
with an additional four satisfying
the condition but for an increase in R and T$_d$ with the observation
by {\it WISE} which we interpret as the signal of the formation of a new
puff or cloud of puffs. In Figure 11, we show the R versus T$_d$ 
relations for seven of the ten stars and also for V854 Cen, a star with
a near-constant R and T$_d$. An evolutionary trend is indicated for
each star. Note that for RT Nor and V CrA, the {\it WISE} result is disconnected from
 the indicated  trend from earlier photometry because, we suppose,  fresh puffs have
 appeared between the time of the {\it Spitzer} and {\it WISE}
observations.  Inspection of Figure 11 shows that the evolutionary
trends comprise a very diverse set. 
This diversity is likely to reflect several factors including the 
variation from star to star of the dust content, the different stellar
luminosities and temperatures across the sample. RCBs may evolve at
approximately constant luminosity but with a finite luminosity range
across the sample. Stellar temperatures (see Paper I) for the stars represented by Figure 11 run from 4200 K
for V1157 Sgr and SV Sge to 8000 K for V3795 Sgr. 

If the simple model leading to the prediction that $f_{dust} \propto$ T$_d^4$ for
a given star is applicable, the diversity exhibited in Figure 11
will be markedly reduced when data for the sample are displayed after
scaling such that R/R$({\it IRAS})$ is plotted versus $T_d/T_d({\it IRAS})$,
as in Figure 12. Moreover, the model predicts that the scaled
R and T$_d$ should be related by $R \propto T_d^4$. This prediction shown in
Figure 12 is a fine fit to the observational results.  

When the other RCBs are added to Figure 12, they generally confirm the predicted
trend but with increased scatter which may be attributed to their cloud of puffs is
changing over the time period covered by the observations. The fact that the majority of RCBs
satisfy the predicted relation suggests that the evolution  of puffs around follows a fairly
common pattern. In three cases -- RT Nor,
GU Sgr and FH Sct -- the R value  and T$_d$  relative to their {\it IRAS} values exceed unity suggesting that
puffs have formed close to the RCB star.  These observations fall between the  R vs T$_d^4$ and T$_d^2$
relations.  

 As pointed in Paper I, it is very puzzling that the {\it IRAS} flux densities (Table 1) and now the R values (Table 2)
are almost without exception greater than all observations since 1983.  The
implication is that almost all RCBs were at their infrared brightest in 1983 when
observed by {\it IRAS}. Two speculations of this surprising
result suggest themselves in isolation or in cooperation: 
(i) there is a systematic difference between the determination of
the parameters $R$ and $T_d$ from the {\it  IRAS} 12 and 25 $\mu$m flux
densities  and the three other satellites sources,
and (ii) the dust shells (or, perhaps, the central stars) have evolved
systematically over the thirty year interval since the {\it IRAS}
observations. An explanation intrinsic to the RCBs is  inconceivable.
(point ii). It 
implies that the 1983 epoch was somehow a critical time of high $R$ and
high $T_d$ for essentially each of the RCBs in our sample. This is
most unlikely given that the RCB phase lasts
several thousand years (Iben 1991).

\begin{figure}
\epsfxsize=8truecm
\epsffile{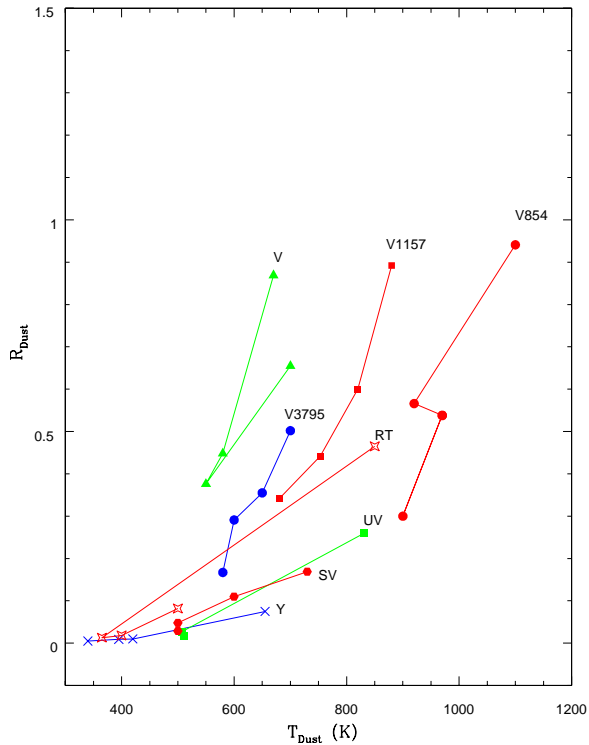}
\caption{Variation of T$_d$ with R for the sample of eight stars identified by the key on the figure. With the exception of V854 Cen, the chosen
stars show a decline in T$_d$ and R from {\it IRAS} to {\it WISE} with two -- V CrA and RT Nor -- showing  an increase in mid-IR
flux with the {\it WISE} observation. } 
\end{figure}

\begin{figure}
\epsfxsize=8truecm
\epsffile{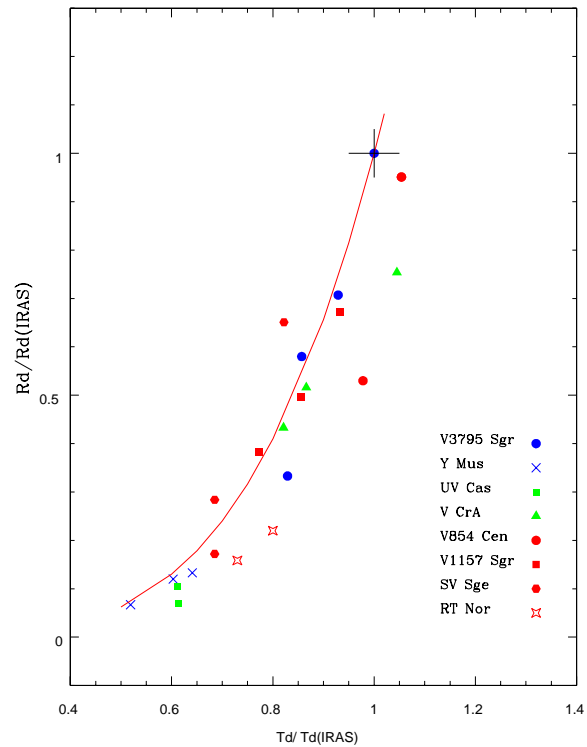}
\caption{Normalized dust temperatures T$_d$/T$_d({\it IRAS})$ and normalized covering factors R/R({\it IRAS}). Points given for the eight stars
represented in Figure 11 fit the predicted relation discussed in the text. } 
\end{figure}


\subsection{Dust and Photospheric Composition}

\begin{figure}
\epsfxsize=8truecm
\epsffile{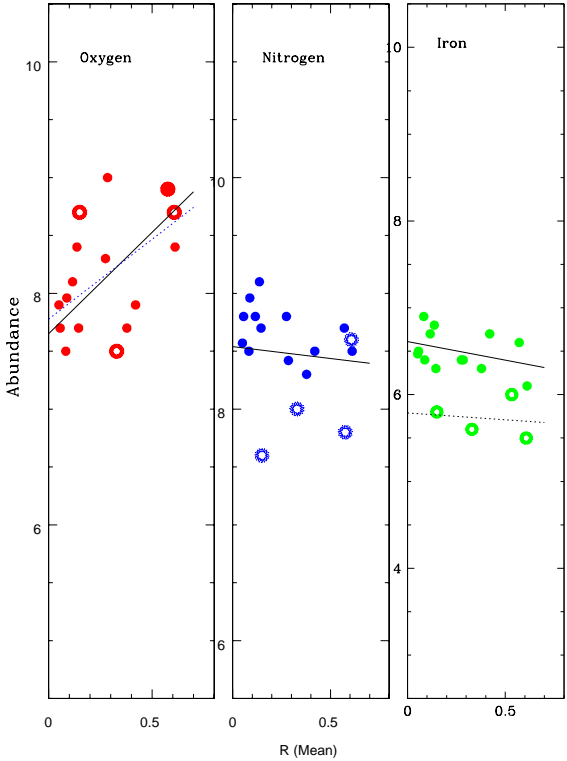}
\caption{ Abundance of photospheric oxygen (red symbols ), nitrogen
 (blue symbols) and iron (green symbols) are plotted with respect
 to the mean R from Table 2. Majority RCBs are represented by filled circles and minority RCBs by open circles.}
\end{figure}

\begin{figure}
\epsfxsize=8truecm
\epsffile{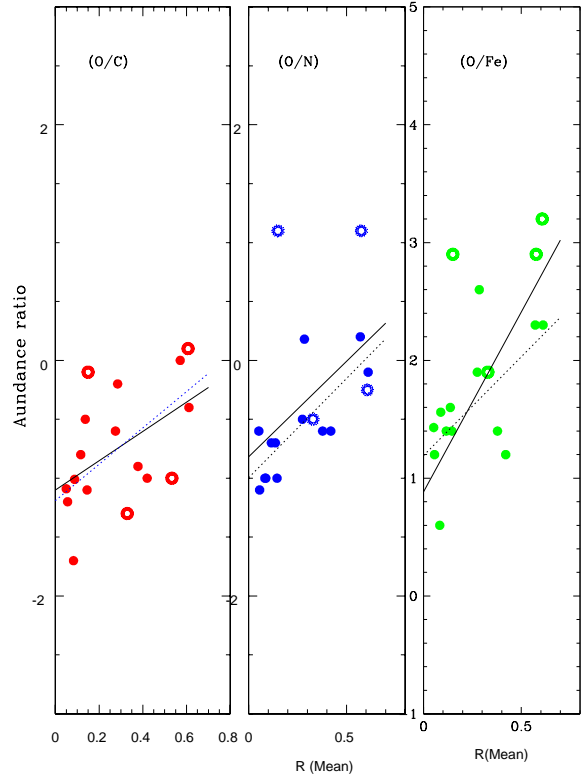}
\caption{Photospheric abundance ratio O/C (left panel), O/N (central panel) and O/Fe (right panel) for majority (filled circles) and minority (open circles) RCBs. The solid line is the least
squares fit  to all stars and the dotted line is the fit for majority RCBs alone.  }
\end{figure}

Photospheric composition should be one of the key variables influencing the formation of the carbon soot, a presumed principal
component of a puff's dust.  Chemical compositions of  warm RCBs come mainly from Asplund et al.  (2000) with results for V854 Cen
from Asplund et al. (1998), V532 Oph from Rao et al. (2014) and V2552 Oph from Rao \& Lambert (2003).   Two introductory points about these
compositions. First,  the analysis by Asplund et al. (2000) using modern H-poor He-rich model atmospheres (Asplund et al. 1997) revealed what
was termed `the carbon problem', i.e.,  the C abundance returned from the analysis of a spectrum was about 0.6 dex less than the C abundance adopted in
the model atmosphere's construction. This problem was discussed by Asplund et al. (2000) who suggested that an abundance ratio was most likely more reliable than the abundance of either element.  Second, Lambert \& Rao (1994)
 in their abundance analysis, a precursor of that reported by Asplund et al. (2000), noted that the warm RCBs could be divided by composition into a
 `majority' and `minority' group. The minority RCBs differed from the majority principally  in having a lower Fe and higher Si abundance and, hence,
 an extraordinary Si/Fe ratio, a ratio far in excess of that seen in any other major class of star such as metal-poor halo stars.   Table 1 includes the
 minority stars V854 Cen, V3795 Sgr, VZ Sgr and V CrA.
 
The mean R -- the quantity  R$_{\rm av}$ from Table 2 --  is taken as a RCB's propensity to shed puffs.  The C abundances -- see above remarks on the carbon
 problem -- span a small range  including both majority and minority RCBs and are uncorrelated with the mean R.  The O
 abundances exhibit a range of about 1.5 dex.  There appears to be a positive correlation
 between the O abundance and the mean $R$ (Figure 13) with barely a separation between majority and minority RCBs. 
 The N abundances of  majority RCB show a weak anti-correlation with the mean $R$ with three of the four minority RCBs having a far lower N abundance
 (Figure 13).  Similarly, the Fe abundances of the majority are weakly anti-correlated with the mean $R$ with, of course, the Fe abundances of minority RCBs falling below those of the majority RCBs.
   A least squares solution of all the stars (both majority and
  minority)  for O abundance  with respect to the mean R suggests a slope
  of 1.24$\pm 0.4$ and a correlation coefficient of 0.52  which would increase
  to 1.33$\pm 0.33$ and a correlation coefficient of 0.70 if two outliers
  are dropped. For the N abundance with respect to the mean R, the  slope is
   -0.5$\pm 0.4$ with a correlation coefficient of -0.3 . For Fe abundance
  the majority RCBs  suggest a slope of -0.4$\pm 0.3$ with a correlation 
  coefficient of -0.37. For minority RCBs the  Fe abundances give  the slope is -0.15$\pm 0.7$
  with a correlation coefficient of -0.14.

Correlations between abundance ratios and the mean R have also been looked for. 
  Using the spectroscopic C abundance, the O/C
 ratio also shows a weak correlation with  mean $R$  without a distinction between majority and minority RCBs.
  A least-squares solution of the majority RCBs shows a slope of 1.54 and a
  correlation coefficient of 0.65. The slope becomes 1.24$\pm 0.4$
  and correlation coefficient of 0.5 when  majority and minority RCBs 
  are  considered together.
   The O/N and O/Fe ratios (Figure 14)  show a rough tendency to increase
   with increasing mean $R$ with two of the four minority.
 RCBs falling amongst the majority RCBs; the exceptions among minority RCBs are VZ Sgr and V854 Cen with O/N and O/Fe ratios greater than those held by 
 any majority RCB.  The least-squares solution for O/N shows a slope 
of 1.7$\pm 0.4$ and correlation coefficient of 0.77 for majority and 0.55 for
 combined sample with the same slope. The O/Fe also suggests a slope of 1.67
 $\pm 0.67$ and a correlation coefficient of 0.60 for majority and combined
 sample has a slope of 2.17$\pm 0.7$ . All three abundance ratios O/C, O/N,and
 O/Fe seem to show a positive slope with respect mean R.  
               
 Thus, it is possible that O abundance might have an influence in dust 
 production in these stars.  Woitke et al.'s (1996) models for dust formation
 in RCBs suggest that the CO molecule is the most dominant cooling agent which
 determines whether the gas can reach condensation temperatures or not
 after passage of a pulsation shock. As long as the C abundance exceeds the O abundance, the CO
 abundance is likely to be influenced by the O abundance.
 It is not quite clear how the CO abundance in these stars is controlled by oxygen abundance.

\subsection{Dust  and Luminosity}

Is dust formation  related to a star's luminosity?  In the absence of  traditional (i.e., trigonometric)  methods of estimating luminosity,  we rely
on the pulsation-period relation for RCBs.  Pulsation is also of direct interest to the puzzle of dust formation in that
a link, first suggested by Pugach (1977) and confirmed by
 Crause et al. (2007) ties onset of dust formation to a preferred phase in a RCBs radial pulsation. However, copious dust formation
 appears not to occur at every passage through the preferred phase. Also, it is unclear how dust formation is physically related to this
 phase.  Although Woitke et al. (1996) developed a model for a  shock-induced dust 
 formation in a pulsating RCB, many details are still to be worked out, particularly in
 applying it to specific stars (however see Rao \& Lambert 2010).
 
  On the theoretical front, radial pulsations obey the relation $P \rho^{1/2} = Q$ 
   where P is the period,  $\rho$ the density and $Q$ is a constant. 
  Following Fadeyev (1996),  P in days can be written as 
 follows
 
 \begin{equation}
P = \left(\frac{T_{\rm eff}{\odot}}{T_{\rm eff}}\right)^3\left(\frac{L }{L_{\odot}}\right)^{3/4}\left(\frac{M }{M_{\odot}}\right)^{-1/2} Q
\end{equation}

where $L$ and $M$ are luminosity and mass, respectively.

\begin{table*}
\centering
\begin{minipage}{180mm}
\caption{\Large Pulsations and dust }
\begin{tabular}{llllll}
\hline
  Star & $T_{\rm eff}$  &Period & R(mean)    & Period Ref. &  Comment\\
  & K & days &  &   &  \\
\hline
 UW Cen & 7500   & 42.79  & 0.38 & Crause et al (2007) &  \\
  Y Mus & 7250   & 35.0   & 0.06 & Marang et al (1990) &  \\
 XX Cam & 7250   & 36     & 0.01 & Weiss et al (1996)  &  \\
 RY Sgr & 7250   & 37.79  & 0.42 & Crause et al (2007) &    \\
 UV Cas & 7250   & 40.16  & 0.08 & Weiss et al (1996)  &    \\
 RT Nor & 7000   & 43     & 0.14 & Lawson et al (1990) &   \\
 VZ Sgr & 7000   & 40$^a$ & 0.15 & Lawson \& Cottrell (1997) \\
 UX Ant & 7000   & 50     & 0.14& Lawson et al (1994) &    \\
 RS Tel & 6750   & 45.8   & 0.28& Lawson et al (1990) &    \\
 RZ Nor & 6750   & 42.4$^b$ &0.57& Lawson et al (1990) &  \\
  R CrB & 6750   & 42.7$^c$ &0.28& Rao et al (1999) &  \\
 V854 Cen&6750   & 43.25  & 0.58 & Crause et al (2007) &    \\
 V532 Oph&6750   & 50     & 0.05 & Clayton et al (2009) &  \\
 SU Tau & 6500  & 45.07   & 0.61 & Weiss et al (1996) &  \\
  V CrA & 6500   & 57.1   & 0.61 & Lawson et al (1990) &  \\
 FH Sct & 6250   & 41     & 0.14& this paper          & \\
  U Aqr & 5000   & 81.3   & 0.31 & Lawson et al (1990) & \\
  Z UMi & 5200   & 130    & 0.74 & Benson et al (1994) &  \\
  S Aps & 4200   & 120$^d$& 0.29 & Kilkenny \& Flanagan (1983) &\\
 NSV11154 &4200  & 114$^f$ & 0.65 & unpublished         &  \\
 FQ Aqr & 8750   & 18     &      & Kilkenny et al (1999)& coolEHe\\
LSIV-14 109& 9500& 25     &      & Lawson et al (1993) & coolEHe\\
 NO Ser & 11750  & 22     &      & Kilkenny et al (1999)& coolEHe\\
V2244 Oph& 12750 & 10     &      & Lawson et al (1993) & coolEHe\\
HD175893 &  5500 & 41.0   & 0.026& Lawson et al (1990) & HdC \\
HD137613 & 5400$^e$& 20     &    & Kilkenny et al (1988) & HdC \\
HD148839 & 6500  & 20     &      & Kilkenny et al (1988) & HdC \\
HD182040 & 5600  & 40     &      & Kilkenny et al (1988) & HdC \\
HD173409 & 6100  & 20     &      & Kilkenny et al (1988) & HdC \\
\hline
\end{tabular}
\\ $^a$ - 47days according to Bateson (1978)
\\$^b$ -  68 days?
\\$^c$ - Photometric period changes 43-53 days according to Fernie \& Lawson (1993), Lawson \& Kilkenny (1997) and Percy et al (2005). We use the period determined
 from radial velocities obtained over several decades (Rao et al. 1999) .
\\$^d$ - see the discussion in Lawson et al (1990).The period changes occur.
 The period quoted here might be a fundamental mode as detected by Waters (1966)
 and Kilkenny \& Flanagan (1983).
\\$^e$ -  Kipper (2002) gives $T_{\rm eff}$ of 6000 K and log g of 1.3 for
HD137613 (= HM Lib).
\\$^f$ - Period is variable. The mean period may be 114 days.
\end{minipage}
\end{table*}

                                                               1414,1        98%

 Pulsation periods for many RCBs have been estimated from optical photometry with measurements available in the
 literature for the majority of stars in Table 1 (see Table 3 and Lawson \& Kilkenny 1996).  Photometric amplitudes are small in most cases: less than several 
 hundredths of a  magnitude except for RY Sgr and a  few others. Amplitudes may be variable and $P$ may vary too. With few exceptions, the period $P$ is in the range  30 -- 50 days.  Exceptions
 include U Aqr with $P=81.3$ days (Lawson et al. 1990), $P$ = 114 days for NSV 11154 (our unpublished estimate), $P$=120 days for S Aps (Kilkenny \& Flanagan  1983) and $P$=130 days for
 Z UMi (Benson et al. 1994).  Periods shorter than 30 days are found among the extreme hellum stars,  the hydrogen-deficient
 carbon stars and the hot RCBs.  The $T_{\rm eff} vs P$ plane is shown in Figure 15. This plot is similar to Lawson et al.'s (1990) figure 32.  Effective temperatures for most RCBs are spectroscopic values from Asplund et al. (2000). However for the cooler RCBs, except for Z UMi, no spectroscopic
$T_{\rm eff}$s are available and $T_{\rm eff}$ estimates are  based on SED fits and photometry.
 Using Fadeyev's formula, we find 
 most of the RCB stars are distributed around  the locus
 $\log L/M = 4.0$ for M of 0.7 ${M_{\odot}}$ with the HdC stars located around
 $\log L/M = 3.5$ .  
 
  Within the large sample with $P$ between 30 and 50 days, there is no
 correlation between the mean $R$ and $P$ or $T_{\rm eff}$.
 Outside this period range, there are too few RCBs to search for
 connections with the mean $R$. However as seen from Table 2, generally cooler
 RCBs (presumably with longer periods) have larger mean $R$ values.
 At periods shorter than about 20 days, dust formation either does not occur or is an infrequent occurrence. As noted above, the stars in this
 category are the EHes warmer than RCBs,  the HdC stars at the cool end and the hot RCBs at the hot end of the range for RCBs. Only one HdC star that showed 
 mild infrared excess, V4512 Sgr, is marked in  Figure 15. 
 
                 Better estimates of both  periods as well as the mode of
 pulsation, particularly for cool RCBs, are required for a proper assessment 
 of the pulsation- dust connection.

\begin{figure}
\epsfxsize=8truecm
\epsffile{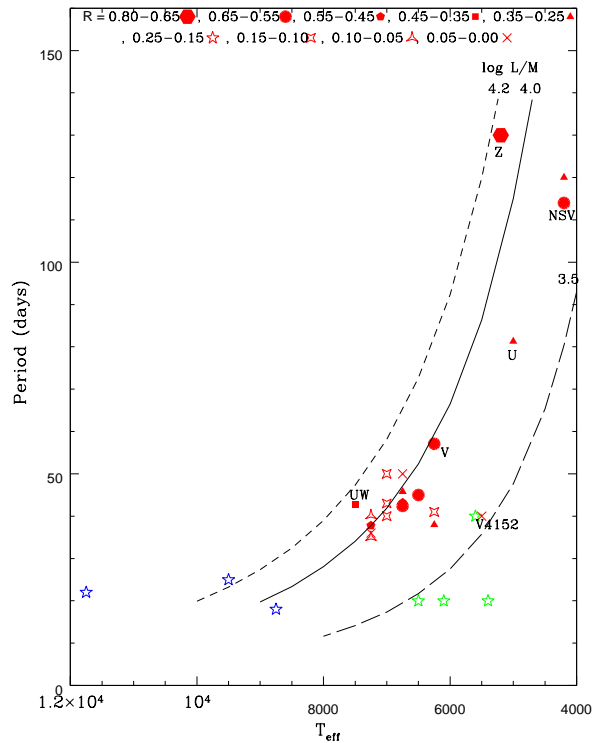}
\caption{Effective temperature $T_{\rm eft}$ versus pulsation period $P$ for RCBs (red symbols), cool EHes (blue symbols) and HdCs (green symbols).
Loci of constant $\log L/M$ are shown for values of 4.2, 4.0 and 3.5 where $L$ is the stellar luminosity and $M$ is the stellar mass. The key for the 
 mean R values is shown on top. Larger mean R is represented with a larger
 symbol. The only HdC star with IR excess, V4512 Sgr (HD175893) is shown with
 a red cross. }
\end{figure}

\subsection{ Dust Mass}

                    It might be relevant to estimate  the mass of dust 
 that is hanging around these stars.  The basic relation is

\begin{equation}
M_d = \left(\frac{F_{\lambda}}{\kappa_{\lambda}}\right)\left(\frac{D^{2}}{B_{\lambda}(T_d)}\right)
\end{equation}

\noindent where $M_d$ is the dust mass, $F_{\lambda}$ is the excess flux
over the stellar flux, $\kappa_{\lambda}$ is the absorption coefficient
of the grain in cm$^{2}$gm$^{-1}$, D is the distance to the object and $B_{\lambda}(T_d)$
is the Plank function .

This quantity also depends on the  estimate of the distance to the object.
 We use the $M_{v}$ vs (V-I) colour established for LMC RCBs by Tisserand et al.
 (2009). As typical for F type stars UV Cas and Y Mus , we assume the $M_{v}$ to be
 -5 and estimate the distance. Also assuming the dust to be amorphous carbon
 of BE type (Colangeli et al. 1995) we estimate the dust mass at the time of IRAS
 observations as 2.9 x 10$^{-9}$ m$\odot$ for both UV Cas and Y Mus. A similar estimate for cool RCB Z UMi is obtained, by assuming $M_{v}$ of -3.5 as consistent with the
 observed V-I colour, as  4.8 x 10$^{-9}$ m$\odot$  at the time of IRAS 
 observations.
     
                Clayton et al.  (2011) imaged R CrB and its environment during the
  prolonged deep minimum that started in 2007. They discovered several
 cometary knots around the star which are interpreted as  past ejected dust
 puffs from the star. As they state `The puffs cause declines when they form
 directly in our line of sight and may be seen as cometary knots when they 
 form to the side of or behind R CrB'. It is interesting to note that the dust mass
 estimated for these knots of 10$^{-8}$ m$\odot$ to  10$^{-9}$ m$\odot$ is about equal
 to dust mass estimated above for UV Cas, Y Mus and Z UMi from midIR excess
 at any given instant.  It is of interest to note that Clayton et al. (2011)
  estimate total dust mass of 10$^{-2}$ m$\odot$  for the R CrB shell
 including the cold dust from their observations into far infrared with
 {\it Herschel}.

\section{Concluding remarks}

Variations in the mid-IR flux densities measured by the  different
satellites reflect changes in the circumstellar shell's mid-IR emission.   Plots comparing
flux densities from the four IR satellites are provided in Figures 1, 2 and 3. It is of interest to
tease from such comparisons the intrinsic changes in
mid-IR emission.
Identification of intrinsic changes is certainly secure for the
extreme outliers in the figures -- most notably, UV Cas  (Figure 4),
S Aps  (Figure 5)  and RT Nor  (Figure 6).  These must arise from the varying
contribution of puffs in the circumstellar shell; the scale of these
variations far exceeds the measurement uncertainties. As discussed
above, UV Cas's high {\it IRAS} flux densities arise from puffs
emitted somewhat prior to the {\it IRAS} observation and, as these puffs
evolved away from the star and cooled, lower flux densities were
measured by the later satellites and in the ground-based important series
of measurements by Bogdanov et al. (2010). Somewhat similarly, RT Nor owes its
position as an outlier  to the appearance of fresh puffs between the
{\it Spitzer} and {\it WISE} observations and some fraction of these puffs
caused the deep optical decline recorded by {\it AAVSO} observers.

Extreme outliers are valuable because they provide a measure of
the contribution to the mid-IR emission
from an individual puff or a collection of associated puffs;  RT Nor experiences a 0.4 increase in
$R$ between the {\it Spitzer} and {\it WISE} observations. Other RCBs
with a circumstellar shell occupied by several to many puffs will
experience smaller variations in their mid-IR emission. One might suppose
that variations will decrease in size as the number of puffs increases.

Stars where the mid-IR emission appears to be dominated by the slow evolution of the
circumstellar dust offer the opportunity to test theoretical ideas. In particular,  movement of  a given family of
puffs away from the star is predicted to lead to the relation R$ \propto T_d^4$. As shown in Section 3.3 and, in particular by
Figure 12, this prediction is well verified by stars where the mid-IR emission is thought to come from  a single collection 
of puffs.  In other cases, the observations scatter about the predicted relation in a plausible fashion.

Puff creation aside, evolution of the mid-IR emission from existing
puffs is anticipated to be a slow process, as revealed by the case of
UV Cas. Equilibrium temperatures of a gray grain (see Paper I and above)
about a typical RCB are predicted to run from about 1300K at 10 stellar
radii to 500K at 50 stellar radii to 250K at 200 stellar radii. If the
dust grains are repelled from the star at about 10 km s$^{-1}$, their
path from their formation at (say) 10 stellar radii to about 200 stellar
radii will take about 40 years. Thus, over the duration spanned by {\it AKARI},
{\it Spitzer} and {\it WISE}, the mid-IR emission from a given set of puffs will  evolve very little,
except for the formation of fresh puffs. Optical depth effects at visible and
infrared wavelengths may complicate this simple expectation. Extending the
time span to include {\it IRAS} should reveal a steady evolution, as indeed is shown in Figure 12.
Toward the end of a deep decline, a high-velocity wind  is seen  as a blue-shifted absorption component to the
Na D and other strong resonance lines: radial velocities attain about 200 km s$^{-1}$ (Rao et al. 1999).  This wind's origin is unknown but,
if it can sweep outward gas and dust ahead to it, it offers an explanation for some examples (e.g., V3795 Sgr) of evolution of mid-IR
emission much more rapid than  observed for UV Cas and similar cases.  Of course, this high-speed wind can affect
the mid-IR emission along any radius vector where puffs have accumulated.   
Evolutionary timescales for RCBs are several hundreds of years at least. One may suppose that a RCB's propensity to form and
eject dusty puffs evolves during this time. Some interesting insight that emerged
 from these midIR studies of RCBs are that the oxygen abundance has some influence
 on the dust production. It also appears that cool RCBs have more dust around
 them and higher luminosity at a given mass helps in dust production. Studies
 of pulsations, particularly in cool RCBs, would be a great help in future understanding of dust production in these stars.

  Observational insights into the formation and evolution of dust around RCB stars are limited by the paucity of measurement at
IR wavelengths.  Campaigns extending Bogdanov et al.'s (2010) close scrutiny of UV Cas to other RCBs are to be
encouraged.  Such a campaign should extend to long wavelengths the valuable two decade long programme of  {\it JHKL}
photometry of a dozen southern RCBs by Feast et al. (1997; see also Feast 1997).  Although, as shown here, the four IR satellites
from {\it IRAS} to {\it WISE} have provided valuable insights into the evolution of circumstellar dust for a major sample of RCBs, it is most
unlikely that a satellite will ever provide a thorough collection of IR photometry of these rare and fascinating stars.

\section{Acknowledgements}
                       We  appreciate various comments made by the referee
 Geoff Clayton which improved the paper a great deal.
   We thank Anibal Garc\'{\i}a-Hern\'{a}ndez for his assistance with the preparation and analysis of the observations of RCBs with
  the {\it Spitzer} satellite.                                                                                   
We acknowledge with thanks the variable star observations from the 
AAVSO  database.
This research has made use of the SIMBAD database, operated
at CDS, Strasbourg, France.
 This research has been supported in part by the grant F-634 from
the Robert A. Welch Foundation of Houston, Texas.



\end{document}